\documentclass[prd,a4paper,showpacs,preprint,byrevtex]{revtex4}
\usepackage{graphicx}
\usepackage{dcolumn}
\usepackage{amsmath}
\usepackage{array}
\usepackage{bm}
\usepackage{amssymb}
\usepackage{latexsym}
\usepackage{slashed}
\usepackage{textcomp}

\newcommand{\ZZ}{\mathbb Z}

\newcommand{\mathun}{\hbox{ 1\hskip -3pt l}}

\newcommand{\be}{\begin{equation}}
\newcommand{\ee}{\end{equation}}

\def\mathun{\hbox{ 1\hskip -3pt l}}
\begin{document}

\title{On bound states of Dirac particles in gravitational fields}

\author{Nicolas \surname{Boulanger}}
\thanks{Postdoctoral FNRS Researcher}
\email[E-mail: ]{nicolas.boulanger@umh.ac.be}
\author{Philippe \surname{Spindel}}
\email[E-mail: ]{philippe.spindel@umh.ac.be}
\affiliation{Service de M\'{e}canique et Gravitation,
Universit\'{e} de Mons-Hainaut,
Acad\'{e}mie universitaire Wallonie-Bruxelles,
Place du Parc 20, BE-7000 Mons, Belgium}
\author{Fabien \surname{Buisseret}}
\thanks{FNRS Research Fellow}
\email[E-mail: ]{fabien.buisseret@umh.ac.be}
\affiliation{Groupe de Physique Nucl\'{e}aire Th\'{e}orique,
Universit\'{e} de Mons-Hainaut,
Acad\'{e}mie universitaire Wallonie-Bruxelles,
Place du Parc 20, BE-7000 Mons, Belgium}

\date{\today}

\begin{abstract}
We investigate the quantum motion of a neutral Dirac particle bouncing on a mirror in 
curved spacetime. We consider different geometries: Rindler, Kasner--Taub and Schwarzschild, and show how to solve the Dirac equation by using geometrical methods. We discuss, in a first-quantized framework, the implementation of appropriate boundary 
conditions. This leads us to consider a Robin boundary condition that gives the quantization of the energy, the existence of bound states
and of critical heights at which the Dirac particle bounces, extending the well-known results established from the Schr\"odinger equation. We also allow for a nonminimal coupling to a weak magnetic field. 
The problem is solved in an analytical way on the Rindler spacetime. 
In the other cases, we compute the energy spectrum up to the first relativistic corrections, 
exhibiting the contributions brought by both the geometry and the spin. 
These calculations are done in two different ways. On the one hand, using a relativistic 
expansion and, on the other hand, with Foldy--Wouthuysen transformations.
Contrary to what is sometimes claimed in the literature, both methods are in agreement, as expected. 
Finally, we make contact with the GRANIT experiment. Relativistic effects and effects that go beyond the equivalence principle escape the sensitivity of such an experiment. However, we show that the influence of a weak magnetic field could lead to observable 
phenomena.
\end{abstract}



\maketitle

\section{Introduction}\label{intro}

Quantum mechanical systems in interaction with an external gravitational field are 
of particular interest, not only in theoretical physics, but also from an experimental 
point of view. Thirty years ago, the COW experiment~\cite{cow} displayed
the quantum mechanical phase shift experienced by neutrons waves due to their interaction 
with the Earth's gravitational field. 
The acceleration of a sodium atom due to gravity has also been measured, see Ref.~\cite{chu}. 
More recently, the GRANIT experiment proved the existence of bound states of ultracold 
neutrons bouncing on a mirror in the Earth's gravitational field~\cite{nesv02}. 
Most of the theoretical studies related to these experiments feature the Schr\"odinger 
equation in the linear Newtonian potential~\cite{nesv02,the1}. 
This simple model gives satisfactory results in reproducing the 
experimental data because the particles or atoms used are nonrelativistic. 
\par The next step in the improvement of such models should be the inclusion of the 
spin degrees of freedom. In particular, the dynamics of neutrons should be described by 
the Dirac equation on curved spacetime. 
Many theoretical investigations have been devoted to the study of the spin--gravity couplings 
of a Dirac particle on the Rindler spacetime (see for example Refs.~\cite{Hehl,Obu,Sil}). 
Few of them focused on the problem of bound states of Dirac particles in a gravitational field (see, however, \cite{Arm}). 
The latter problem, which we investigate in the present work, is motivated by the 
GRANIT experiment but also has an intrinsic theoretical interest related to the 
understanding of gravitational effects in first-quantized systems beyond the 
equivalence principle \cite{hawk}.  
\par We focus on a neutral Dirac particle bouncing on a fixed mirror in a gravitational 
field. Three particular geometries are considered. The first one involves the Rindler metric, 
which corresponds to the metric of a uniformly accelerated observer~\cite{rind}. 
It is generally believed to describe a homogeneous gravitational field. 
The second metric with a cylindrical symmetry that we use is the Kasner--Taub metric, which is a particular case of Kasner metrics~\cite{Kas, Taub}. 
In contrast with the Rindler metric, it describes a genuinely curved spacetime, 
corresponding to the spacetime in the neighborhood of a plane brane~\cite{Ph}, 
see also~\cite{Bed}. 
Because of the equivalence principle, both Rindler and Kasner--Taub metrics are equivalent 
in the Newtonian limit, but differ in relativistic corrections as we show. 
Finally, we also use the well-known Schwarzschild geometry. 

Our goal is to compute the energy  spectrum of the Dirac particle. 
The energy is indeed quantized because of the mirror which imposes particular boundary 
conditions. 
Using the Rindler metric, we can compute the energy spectrum in an analytical way. 
For the other geometries, we restrict ourselves to the first relativistic corrections 
that we compute by resorting to the usual relativistic expansion and by using a 
Foldy--Wouthuysen (FW) transformation~\cite{FW}. 
It is of interest to compare these two different methods since the validity of the 
FW transformation in the present context is still subject to
discussions, see Ref.~\cite{Sil} and references therein. 
We also allow for the neutral Dirac particle to be nonminimally coupled to a weak magnetic field and show that some observable effects could be derived in 
future experiments. 
\par Our paper is organized as follows. 
In Section~\ref{Dform} we introduce the notation, write the Dirac equation on
curved spacetime and focus on cylindrically-symmetric geometries. 
We also discuss the boundary conditions encoding at best the presence of the mirror on 
which the particle bounces. In Section~\ref{Dsphe} we proceed to the same analysis  
with a spherically-symmetric geometry. 
In Section~\ref{metric} we particularize the metrics that we use: Rindler and 
Kasner--Taub. Then, we solve the Dirac equation on those 
spacetimes. 
On the Rindler spacetime, the Dirac equation can be solved analytically, taking into 
account the boundary condition induced by the mirror.  
Moreover, we compute the energy spectrum of the particle in the Rindler and Kasner--Taub
spacetimes using a relativistic expansion.  
In Section~\ref{Pauli}, the same computations are done, but this time by resorting to a FW transformation and also considering the Schwarzschild geometry.  
Finally, we apply our results to neutrons in the Earth's gravitational field and 
compare them to the GRANIT experiment in Section~\ref{compar}. 
Our conclusions are given in Section~\ref{Conclu}.         

\section{Dirac equation on a cylindrically-symmetric spacetime}  
\label{Dform}

\subsection{Dirac equation in curved spacetime} 
We introduce the notation and recall the main theoretical features that we 
will need about the Dirac equation on curved spacetime. 
First of all, the Clifford algebra is given by $\{\gamma_{a},\gamma_{b}\}=2\eta_{ab}$ where 
$a,b=0,\ldots,3\,$. The metric is given by $ds^2=\eta_{ab}\, \theta^a\otimes\theta^b$, where 
$\theta^a$ are the coframe fields that diagonalize the metric. 
We adopt the ``mostly minus'' convention 
$\eta= $\;diag$(+---)$ and use units such that $\hbar=1=c$.
 
The covariant derivative for a two-spinor in a curved background has been known for  
a long time \cite{Weyl:1929}. When acting on a Dirac spinor $\Psi$, it is 
given by
\begin{eqnarray}
	\nabla\Psi &=& d\Psi - \frac{i}{2}\,\omega^{ab}\,\Sigma_{ab}\Psi
 = (d + \Omega) \Psi\,, \qquad
  \Sigma_{ab} = \frac{i}{4}{[}\gamma_a,\gamma_b{]}\,,
\nonumber
\end{eqnarray}
where $\Sigma_{ab}$ are the generators of Lorentz transformations and $\omega^{ab}$ are 
the usual Levi-Civita connection one-forms. 
An important property of $\nabla$ is that it preserves the Dirac 
matrices, \textit{i.e.} 
\begin{eqnarray}
\nabla \gamma_b = d \gamma_b - \omega^{c}_{~\,b}\,\gamma_c + [\Omega,\gamma_b] \equiv 0\,.
\nonumber
\end{eqnarray}
\par Finally, the Dirac equation on a curved background reads 
\begin{eqnarray}
	i\gamma^{a}e^{\mu}_{~a}\nabla_{\mu}\Psi -m \Psi =
i	\gamma^a\nabla_{a}\Psi - m \Psi = 0\, ,
	\label{DiracEqn} 
\end{eqnarray}
where the symbols $e^\mu_{~a}$ denote the coefficients of the vierbeins $e_a$ 
in coordinates. 
In the following, Latin indices will denote orthonormal indices, 
whereas Greek indices will denote curved (holonomic) indices. 
Hated specific indices will also be used to denote flat (non-holonomic) indices.   
%
\subsection{Explicit form} 
%
The metric
\begin{equation}
	ds^2=A^2(\zeta) dt^2-B^2(\zeta) (dx^2+dy^2)-d\zeta^2
\label{met_ansa}
\end{equation}
clearly describes a spacetime with cylindrical symmetry along the $\zeta$ axis. 
For the moment, the functions $A$ and $B$ are arbitrary functions of $\zeta\,$; 
they will be specified later on.
Knowing the metric, it is straightforward to write the (co)frame fields:
\begin{eqnarray}
	\left\{ \begin{array}{cc}
	e_{\hat0} = & A^{-1}\partial_t \\
	e_{\hat{\jmath}} = & B^{-1}\partial_j \\
	e_{\hat3} = & \partial_{\zeta}
	\end{array}\right. \quad \,,\qquad
	\left\{
	\begin{array}{cc}
	\theta^{\hat0} = & A \,dt \\
	\theta^{\hat\jmath} = & B\, dx^j \\
	\theta^{\hat3} = & d\zeta
	\end{array}\right. \,,\qquad  j=1,2\,.
	\nonumber
\end{eqnarray}
{}From the zero-torsion condition, we deduce the only 
nonzero components of the spin connection one-form ${\omega}^a_{~b}\,$:
\begin{eqnarray}
	\omega^{{\hat0}}_{~\,{\hat3}} = \frac{A'}{A}\;\theta^{{\hat0}}\,,
	\qquad
	\omega^{\hat\jmath}_{~\,{\hat3}} = \frac{B'}{B}\;\theta^{\hat\jmath}\,,
	\nonumber
\end{eqnarray}    
the prime denoting a derivative with respect to the variable $\zeta\,$.
\par We want to study a neutral fermion of spin $1/2$ with mass $m$ and anomalous magnetic 
moment $\mu_n$, not only in a gravitational field, but also in a weak external electromagnetic 
field described by the field strength $F^{\mu\nu}$. Including such an electromagnetic 
interaction is achieved by adding to the Dirac equation a nonminimal coupling term of the form 
$-\mu_n\Sigma_{ab}F^{ab}$ \cite{Pauli}. If the electromagnetic field is purely magnetic, this 
term reduces to $\mu_n\Sigma_{\hat{k}}{\cal B}^{\hat{k}}$, with 
${\cal B}^{\hat{k}}$ the magnetic field ($k=1,2,3$) and $\Sigma^{\hat{k}}$ the spin matrices.
Using the previous results, we have explicitly from Eq.~(\ref{DiracEqn})
\begin{eqnarray}
	\Big[i\gamma^{\hat0} A^{-1}\partial_t + i\gamma^{\hat\jmath} B^{-1}\partial_{j} +
	i \gamma^{\hat3}\partial_{\zeta} + \frac{i}{2}\,\gamma^{\hat3}
	(\frac{A'}{A} + \frac{2B'}{B}) - m +\mu_n\Sigma_{\hat{k}}{\cal B}^{\hat{k}}\Big]\Psi = 0\,.
	\label{Direxpl}  
\end{eqnarray}
Concerning the metric~(\ref{met_ansa}), it is worth mentioning that we will consider it independent of the magnetic field. The problem of writing cylindrically symmetric solutions 
of the Einstein--Maxwell equations is a nontrivial one, out of the scope of the present work.  
We assume here that both the gravitational and magnetic fields are weak enough so that
one can neglect their mutual interactions.

\subsection{Dirac wave function}
\label{sec:DWF}
%
We are interested in solutions described by the following positive-energy ansatz: 
\begin{eqnarray}
	\psi={\rm e}^{-i\omega t}{\rm e}^{i\vec{k}\cdot \vec{x}}A^{-1/2}B^{-1}\, \Theta(\zeta)\,,
	\quad \omega > 0\,.
\nonumber
\end{eqnarray}
Equation (\ref{Direxpl}) becomes
\begin{equation}\label{Dirstep1}
	\omega\gamma^{\hat0} A^{-1}\Theta-\slashed{k}B^{-1}\Theta
	+i\gamma^{\hat3} \Theta'-m\Theta+\mu_n\Sigma_{\hat{k}}{\cal B}^{\hat{k}}\Theta=0
\end{equation}
where $\vec k=(k^1,k^2,0)$ and $\slashed{k}=k^j\gamma_{\hat\jmath}$ 
with $j=1,2\,$.

We now define the spinor $U_{\sigma}$ such that
\begin{eqnarray}
U^\dagger_{\sigma}U_{\sigma}=1\,,
\quad
\gamma^{\hat0} U_{\sigma}= U_{\sigma}\,,
\quad
\hat{\slashed{k}}\gamma^{\hat3}U_{\sigma}=i\sigma U_{\sigma}\,,	
\label{u_prop}
\end{eqnarray}
with $\hat{\slashed{k}}=\slashed{k}/k$, $k=|\vec k|\,$, and $\sigma=\pm1\,$. 
In general, $U_{\sigma}$ defined by the relations~(\ref{u_prop}) is not an eigenvector 
of the magnetic term $\Sigma_{\hat{k}}{\cal B}^{\hat{k}}\,$. 
It will be such if 
\begin{eqnarray}
	\left[\hat{\slashed{k}}\gamma^{\hat3},\Sigma_{\hat{k}}{\cal B}^{\hat{k}}\right]=0\,,
\nonumber
\end{eqnarray}
or equivalently if $\vec{\cal B}$ is orthogonal to $\vec k$, with ${\cal B}^{\hat{3}}=0\,$. 
Thus, if $\vec k=(k,0,0)$, we will assume that $\vec{\cal B}=(0,{\cal B},0)$. 
Then, the following relations also hold
\begin{eqnarray}
	\slashed{k}U_{\sigma}&=&ik\sigma \gamma^{\hat3}U_{\sigma}\,,
	\nonumber \\
	\Sigma_{\hat{\jmath}}{\cal B}^{\hat{\jmath}}U_{\sigma}
	&=&\frac{{\cal B}\sigma}{2}U_{\sigma}\,.
	\nonumber
\end{eqnarray}
As $U^\dagger\gamma^{\hat3}U=0\,$, it is convenient to decompose $\Theta$ as 
follows:  
\begin{equation}\label{th_def}
	\Theta=F(\zeta)U+G(\zeta) \gamma^{\hat3} U\,,
\end{equation}
where we dropped the dependence in $\sigma$ in order to simplify the notation. 
The normalization condition
\begin{equation}\label{norm0}
	\int \Theta^\dagger\Theta\, d\zeta=1
\end{equation}
implies that the following relation must hold:
\begin{equation}\label{norm}
		\int \left(|F|^2+|G|^2\right)\, d\zeta=1\,.
\end{equation}
Moreover, the application of the condition~(\ref{norm0}) to Eq.~(\ref{Dirstep1}) allows us 
to write the energy of the neutron as 
\begin{equation}\label{energy}
	\omega=\int \Theta^\dagger\hat{H}\gamma^{\hat0}\Theta\, d\zeta\,,
\end{equation}
with
\begin{eqnarray}
	\hat H=-AB^{-1}\slashed{k}+iA\gamma^{\hat3}\partial_\zeta
	+mA\mathun+\mu_n\Sigma_{\hat{k}}{\cal B}^{\hat{k}}.
\nonumber
\end{eqnarray}
In principle, Eqs.~(\ref{norm}), (\ref{energy}) and a boundary condition (discussed below) are sufficient to compute the energy spectrum, provided we know the functions 
$F$ and $G\,$. We thus have to find the differential equations that these functions satisfy. 

Due to the decomposition (\ref{th_def}) and to the fact that the spinors $U$ and 
$\gamma^{\hat{3}} U$ are linearly independent, Eq.~(\ref{Dirstep1}) splits  
into two parts:
\begin{equation}\label{Dirstep2}
\left\{	
\begin{array}{ccc}
D_-\, F-i\varphi\, G-i G'&=&0\,,\\
 -D_+\, G-i\varphi\, F+i F'&=&0\,,
\end{array} 
 \right.
\end{equation}
with 
\begin{eqnarray}\label{D_def}
	D_\pm=\omega A^{-1}\pm m\mp\frac{\mu_n{\cal B}\sigma}{2}\,,\quad \varphi=B^{-1}k \sigma\,.
\end{eqnarray} 
Once $F$ is known, $G$ is readily given by
\begin{equation}\label{Gdef}
	G=\frac{i}{D_+}(\partial_\zeta-\varphi)F\,.
\end{equation}
{}From Eqs.~(\ref{Dirstep2}), one can extract the following equation
\begin{equation}
\label{Dirstep3}
	F''-\frac{D'_+}{D_+}\, F'+\left[D_+ D_--\varphi^2+ \frac{D'_+}{D_+}\, 
	\varphi-\varphi'\right]F=0\,.
\end{equation}
Finally, upon rescaling $F$ according to  
\begin{equation}\label{Fdef}
	F=D^{1/2}_+\, f(\zeta)\,,
\end{equation}
we are left with
\begin{equation}\label{Dirstep4}
	f''+\left[D_+ D_- -\varphi^2+ \frac{D'_+}{D_+}\varphi-\varphi'+\frac{D''_+}{2D_+}-\frac{3D'^2_+}{4D^2_+}\right]f=0\,.
\end{equation}
Together with Eqs.~(\ref{Gdef}) and (\ref{Fdef}), the above equation enables us 
to find $F$ and $G$ once $f$ is known.  
%
\subsection{Boundary condition}
%
The particle we are considering bounces on a mirror placed at $\zeta=0\,$. 
This situation should provide us with a particular boundary condition and 
should eventually lead to the quantization of the energy. 
By analogy with the standard Schr\"odinger equation, one could demand that the wave function 
vanish at the mirror. This condition can easily be implemented with a Klein--Gordon field~\cite{Par} and indeed leads to the existence of bound states~\cite{KG2}. 
However, for a Dirac particle, things are not so simple. For example, imposing  $\left.\Theta\right|_{\zeta=0}=0$ is not very fruitful since the relation (\ref{Gdef}) 
would lead to the conclusion that $\Theta$ is zero everywhere. 
The same conclusion holds if one requires that the 
probability density $j^{\hat{0}}=\Theta^\dagger \Theta=|F|^2+|G|^2$ vanish at the boundary. 
Moreover, the $j^{\hat{3}}$-component of the current is given by
\begin{eqnarray}
	j^{\hat{3}}=\bar\Theta\gamma^{\hat{3}}\Theta=\frac{i}{D_+}(F\, F'^*-F^*F')\,.
\nonumber
\end{eqnarray}
Since $F$ and $F'$ are determined only up to an arbitrary complex coefficient and thus
can be chosen to be real functions (see Eq.~(\ref{Dirstep3})), 
$j^{{\hat{3}}}$ is identically zero. Physically, this expresses the fact that a 
repeatedly bouncing particle is described by a stationary state. 

A satisfactory first-quantized boundary condition can be found if we represent 
the mirror by a scalar potential \cite{fn2} 
$V$ such that $\left.V\right|_{\zeta\leqslant 0}=V_0\gg m$ 
and $\left.V\right|_{\zeta>0}=0\,$. 
Formally, at the level of the Lagrangian, the introduction of such a potential 
amounts to replacing $m$ by $m+V\,$. 
Then, in the region $\zeta<0$ where the potential blows up, Eq.~(\ref{Dirstep4}) becomes approximately
\begin{eqnarray}
	f''\approx V^2_0 f\,,
\nonumber
\end{eqnarray}
which yields
\begin{equation}\label{condi}
	F\approx N\, V^{1/2}_0 {\rm e}^{V_0\zeta}\,,\quad 
	G\approx i N\, V^{1/2}_0 {\rm e}^{V_0\zeta}\, ,
\end{equation}
where $N$ is a normalization constant. 
Since, at $\zeta=0$, the wave function has to be continuous, we assume from  
Eq.~(\ref{condi}) the Robin boundary condition
\begin{equation}\label{quant0}
	\left.F\right|_{\zeta=0}=-i\, \left.G\right|_{\zeta=0}
\end{equation}
which, using Eq.~(\ref{Gdef}), can be rewritten as
\begin{equation}\label{quant}
	(D_++\varphi)\left.F\right|_{\zeta=0}=\left.F'\right|_{\zeta=0}\,.
\end{equation}
Thanks to this last relation, the energy of the neutron will be quantized. The presence of $\varphi$ ensures that only two solutions exist: The energy will be spin-dependent. 
Moreover, at leading order, Eq.~(\ref{quant}) becomes 
$\left.F\right|_{\zeta=0}\approx 0\,$. 
This is the boundary condition we are used to see when the Schr\"odinger and/or the Klein--Gordon equations are employed~\cite{KG2,Flu}.

\section{Dirac equation on a spherically-symmetric spacetime}  \label{Dsphe}

We now turn our attention to spherically-symmetric metrics. Such a metric reads 
\begin{equation}\label{met_sphe}
ds^2= \text{e}^{2\Phi(r)}dt^2-\text{e}^{2\Lambda(r)}dr^2-r^2d\omega^2	
\end{equation}
and the corresponding Dirac equation is
\begin{equation}\label{Dir_sch}
	\left[i \text{e}^{-\Phi}\gamma^{\hat{0}}\partial_t+
	i\gamma^{\hat{3}}\text{e}^{-\Lambda}\left(\partial_r+	\frac{\Phi'}{2}\,+\frac{1}{r}\,\right)
	+\frac{i}{r}\,\gamma^{\hat1}\left(\partial_\theta+\frac{\cot\theta}{2}\right)
	+\frac{i\gamma^{\hat2}}{r\sin\theta}\,\partial_\phi-m\right]\Psi=0\,,
\end{equation}
where a prime now denotes a derivative with respect to the radial coordinate $r\,$. 
It is convenient to use the following representation of the Dirac matrices
\begin{eqnarray}
\gamma^{\hat{0}}=\left(\begin{array}{cc}\mathun&0\\0&-\mathun\end{array}\right)
\qquad,\qquad\gamma^{ \hat{k}}=-i\,\left(\begin{array}{cc}0&\sigma^k\\ \sigma^k& 0
\end{array}
\right)\,.	
\nonumber
\end{eqnarray}
Decomposing the Dirac spinor into a pair of bi-spinors
\begin{eqnarray}
	\Psi=\left(\begin{array}{r}\chi\\\eta\end{array}\right),
\nonumber
\end{eqnarray}
we find that Eq.~(\ref{Dir_sch}) yields the following two equations
\begin{eqnarray}
i\,\text{e}^{-\Phi}\partial_{t}\chi+\sigma^{3}\,\text{e}^{-\Lambda}\,\partial_{r}\eta+\frac1r\slashed{D}\eta+\frac 12\text{e}^{-\Lambda}
\left( \Phi^\prime+\frac 2r\right)\sigma^{3}\,\eta=m\, \chi\,,
\label{eqbs1}\\
-i\,\text{e}^{-\Phi}\partial_{t}\eta+\sigma^{3}\,\text{e}^{-\Lambda}\,\partial_{r}\chi+\frac1r\slashed{D}\chi+\frac 12\text{e}^{-\Lambda}
\left( \Phi^\prime+\frac 2r\right)\sigma^{3}\,\chi
= m\, \eta\,,
\label{eqbs2}
\end{eqnarray}
where $\slashed{D}$ is the Dirac operator on the unit sphere,
\begin{eqnarray}
	\slashed{D}=\sigma^{1}\,\partial_{\theta}+\frac 1{\sin\theta}\,\sigma^{2}\,\partial_{\varphi}+\frac 1{2}\cot\theta\,\sigma^{1}\,.
\label{Dir2sph}
\end{eqnarray}

In order to solve these equations, we may use the well-known decomposition of 
bi-spinors in terms of spherical spinorial harmonics $\Omega$, presented from a 
group-theoretical point of view in (almost) all text books on relativistic quantum mechanics 
(see for example Ref.~\cite{LL}). 
Nevertheless, for pedagogical purposes, we find it useful to reconsider this decomposition 
from a more geometrical point of view, the method having the advantage of being easily 
extended to more general situations (higher dimensions, other geometries, \textit{etc.}). 
Let us separate the variables and write the bi-spinors $\chi$ and $\varphi$ as products 
of the positive-energy exponential ${\rm e}^{-i\omega t}$ multiplied by radial 
functions and spherical spinorial harmonics:
\begin{equation}
\chi={\rm e}^{-i\omega t} F(r)\Omega(\theta,\varphi),\quad \eta 
= -i{\rm e}^{-i\omega t}G(r)\widetilde\Omega(\theta,\varphi),
\label{anssphe}
\end{equation}
where $\widetilde{\Omega}$ has to be specified appropriately. 
By inserting this ansatz (\ref{anssphe}) into Eqs.~(\ref{eqbs1}) and (\ref{eqbs2}), 
it is easy to see that the Dirac equation is satisfied provided \\
(i) we define $\Omega=\sigma^{3}\,\Omega\label{OtO}$\ ;\\
(ii) the spinor $\Omega$ verifies the equation 
\begin{equation}\label{s3D}
	\sigma^{3}\slashed{D} \,\Omega =\kappa \,\Omega\,.
\end{equation}
By integrating $\Omega^\dagger \sigma^{3}\,\slashed{D}\Omega$ by parts on the sphere, it
can be found that the eigenvalues $\kappa$ are real numbers.
On the other hand, we know that the eigenvalues $\mu$ of the Dirac operator~(\ref{Dir2sph}), 
\begin{eqnarray}
\slashed{D}\Upsilon= \mu\,\Upsilon\,,
\nonumber
\end{eqnarray}
are purely imaginary. 
{}From an eigenspinor of the Dirac operator we obtain a solution of Eq.~(\ref{s3D}) 
by taking the combination $\Omega=\Upsilon+i\,\sigma^{3}\,\Upsilon$, the corresponding 
eigenvalues being related by $\kappa=-i\,\mu\,$. 
\par For completeness, let us recall that the eigenspinors of the Dirac operator 
(on any sphere of arbitrary dimension) can be obtained by using a Killing spinor and the 
usual spherical scalar harmonics $Y$ as follows. 
A Killing spinor $K$ is a spinor verifying the equations 
\begin{eqnarray}
D_{\hat a}K_k= \pm  \frac i 2\, \sigma _{\hat a}\,K_k\,.
\nonumber
\end{eqnarray}
On $S^{2}$ there exist four Killing spinors, $k=1,\ldots,4$~\cite{spinor}. 
Taking anyone of them, we may obtain the eigenspinors of the Dirac operator by 
considering the combination 
$\Upsilon_{k}=\partial_{\hat a}Y\,\sigma^{\hat a}\,K_k+ \lambda\,Y\,K_k\,$~\cite{MRooman}. 
Substituting this ansatz into the Dirac equation and using the well-known spectrum of the spherical scalar harmonics, we obtain $\lambda =i\,\mu$ with $\mu\in \ZZ_{0}\,$.

\par The well-known equations that the radial functions $F(r)$ and $G(r)$ have to satisfy are
\begin{eqnarray}
\left\{
\begin{array}{ccc}
{\cal D}_-F-iG'-i\varphi_- G &=& 0\,,\\
-{\cal D}_+ G+iF'+i\varphi_+ F &=& 0\,,
\end{array}\right.
\label{eqsphe}
\end{eqnarray}
where
\begin{eqnarray}
{\cal D}_\pm=\omega\,{\rm e}^{\Lambda-\Phi}\pm m{\rm e}^{\Lambda}\,,
\qquad
\varphi_\pm=\frac{\Phi'}{2}+\frac{1\pm\kappa{\rm e}^{\Lambda}}{r}\;.
\nonumber
\end{eqnarray}
These equations can be seen as the counterpart of Eqs.~(\ref{Dirstep2}). 
{}From Eqs.~(\ref{eqsphe}), we obtain that $F$ and $G$ are given by
\begin{equation}\label{eqsph2}	
F''-\left(\frac{{\cal D}'_+}{{\cal D}_+}-\varphi_+-\varphi_-\right)F'+\left({\cal D}_
+{\cal D}_-+\varphi'_+-\frac{\varphi_+{\cal D}'_+}{{\cal D}_+}
+\varphi_+\varphi_-\right)F=0\,,
\end{equation}
\begin{equation}\label{Gphe}
	G=\frac{i}{{\cal D}_+}\left(\partial_r+\varphi_+\right)F\,,
\end{equation}
these equations being similar to Eqs.~(\ref{Dirstep3}) and (\ref{Gdef}) 
obtained previously in the cylindrically-symmetric case. 

The number $\kappa$ is a nonzero integer which can directly be interpreted in the 
Schwarzschild geometry. 
It represents the eigenvalues of the operator constructed on the time translation Killing vector and on the Penrose--Floyd tensor~\cite{Carter:1979fe}, see also Eq.~(23) of 
Ref.~\cite{Carter:2006hj}. 

Due to the symmetry of the problem, the boundary condition (\ref{quant0}) will be slightly modified. Let us indeed consider that the scalar potential is equal to a large constant $V_0$ in the region $r<R$, with $R$ an arbitrary radius, and vanishes for $r>R$. 
Then, we find from Eqs.~(\ref{eqsphe}) that 
\begin{equation}
	\left.F\right|_{r=R}=-i\, {\rm e}^{\Lambda}\, \left.G\right|_{r=R}\,.
\end{equation}
In the Schwarzschild geometry for example, we have
	\begin{equation}\label{badq}
	\left.F\right|_{r= R}=-\frac{i}{\sqrt{1-\frac{r_s}{R}}}\left.G\right|_{r= R}\,.
\end{equation}
Note that, obviously, such a boundary condition is only valid in the region $R\gg r_s\,$. 
We stress that the present first-quantized formalism is no longer valid near the 
Schwarzschild radius, where a second-quantized formalism is definitely needed. 

\section{Exact solutions}\label{metric}

\subsection{Geometrical preliminaries}

Before proceeding to the resolution of our equations, 
we have to specify the metrics we will use. 
Usually, the flat Rindler metric is used to describe the physics of a uniformly 
accelerated system in flat spacetime (see e.g. \cite{Misner:1974qy} chapter 6) 
\begin{equation}\label{Rind}
	ds^2= (1+g\zeta)^2 dt^2-dx^2-dy^2-d\zeta^2\,.
\end{equation}
With the notation introduced in Eq.~(\ref{met_ansa}), it corresponds to  
\begin{equation}\label{defR}
	A=(1+g\zeta)\,,\quad B=1\,.
\end{equation}
Another spacetime metric, that describes a cylindrically symmetric solution of the vacuum Einstein equations, is given by the Kasner--Taub metric \cite{Taub} 
\begin{eqnarray}
	ds^2 &=& ({z}/{z_0})^{-2/3}\;dt^2 -({z}/{z_0})^{4/3}\;(dx^2+dy^2)-dz^2\,.
\label{met} 
\end{eqnarray}
It can be interpreted as the gravitational field around an infinite plane~\cite{Ph}. 
The constant $z_0$ can be related to the gravitational acceleration in the Newtonian limit. In this limit, we set $z=z_0-\zeta$ and expand the metric component 
$g_{00}$ to first order in $\zeta/z_0$, which gives  
$g_{00}=1+\frac{2}{3}\frac{\zeta}{z_0}= 1+2\phi_N\,$. 
The Newtonian potential is thus $\phi_N = g\zeta$ where $g=\frac 13\,z_0\,$. 
Considering the full relativistic metric~(\ref{met}) 
\begin{eqnarray}\label{met2} 
	ds^2 &=& (1-3\, g\, \zeta)^{-2/3}\;dt^2 -(1-3\, g\, \zeta)^{4/3}\;(dx^2+dy^2)
	- d{\zeta}^2\,
\end{eqnarray}
corresponds to taking
\begin{equation}\label{defK}
	A=(1-3g\zeta)^{-1/3}\,,\quad B=(1-3g\zeta)^{2/3}\,
\end{equation}
in Eq.~(\ref{met_ansa}).
In order to restore the $c$ factors, we have to replace $g$ by $g/c^2$ and $t$ by $ct$ as usual. Let us note that the metrics defined through 
Eqs.~(\ref{defR}) and (\ref{defK}) differ at order $c^{-2}\,$. 
For the sake of simplicity, we shall discuss both simultaneously using
\begin{equation}
\label{defc2}
	A = 1 + \frac{g\zeta}{c^2}\, + (1-a)\frac{g^2\zeta^2}{c^4}\,\;,
	\quad B = 1 - b\frac{g\zeta}{c^2}\,\,,\quad 
  \rm{with}  \quad
	a=-1,\ 1,\quad b=2\,,\ 0\,, 
\end{equation}
for the Kasner--Taub and Rindler metrics, respectively. 

\subsection{Bound states in Rindler metric}\label{BsKR}

Exact solutions of the free Dirac equation on Rindler spacetime 
are well-known (see for example Ref.~\cite{Soffel}, or sometimes in a disguised form, Ref.~\cite{Iyer} and Refs. therein). 
In the present section we particularize these solutions to our problem, 
namely a particle bouncing on a mirror, 
and show that boundary condition (\ref{quant0}) implies the quantization of energy.
 
Instead of the spinor $U_{\sigma}$ of Section~\ref{sec:DWF}, 
we introduce the spinor $W_\sigma$ obeying 
\begin{eqnarray}
	\gamma^{\hat0}\gamma^{\hat3}W_\sigma&=&W_\sigma\,,
	\nonumber \\
	\gamma^{\hat1}\gamma^{\hat2}W_\sigma&=&i\sigma W_\sigma\,.
\nonumber
\end{eqnarray}
The positive-energy Dirac spinor is now defined as
\begin{eqnarray}
	\Psi={\rm e}^{-i\omega t}{\rm e}^{i\vec{k}\cdot \vec{x}}\, \Theta_\sigma(\rho)\,
\nonumber
\end{eqnarray}    
where $\rho = \zeta+\frac{1}{g}\,$ and
\begin{eqnarray}
	\Theta_{\sigma}(\rho)= \lambda^-(\rho) W_\sigma+\lambda^+(\rho) \gamma^{\hat0} W_\sigma\,.
\nonumber
\end{eqnarray}
It can be shown that the functions $\lambda^\pm$ obey the following equations~\cite{Soffel}
\begin{eqnarray}
	\left[\rho\partial_\rho \rho\partial_\rho\right]\lambda^\pm=\left[(m^2+k^2)\rho^2+
	\left(\frac{i\omega}{g}\pm\frac{1}{2}\right)^2\right]\lambda^\pm\,, 
\nonumber
\end{eqnarray}
whose solutions are
\begin{eqnarray}
	\lambda^\pm=H^{(1)}_{i\omega/g \pm 1/2}(i\kappa \rho)\,,
\nonumber
\end{eqnarray}
where $H^{(1)}$ is the Hankel function of the first kind (the Hankel 
function that vanishes when $\rho$ is going to infinity) and $\kappa^2=m^2+k^2\,$.
\par
\begin{figure}[b]
\includegraphics*[width=9cm]{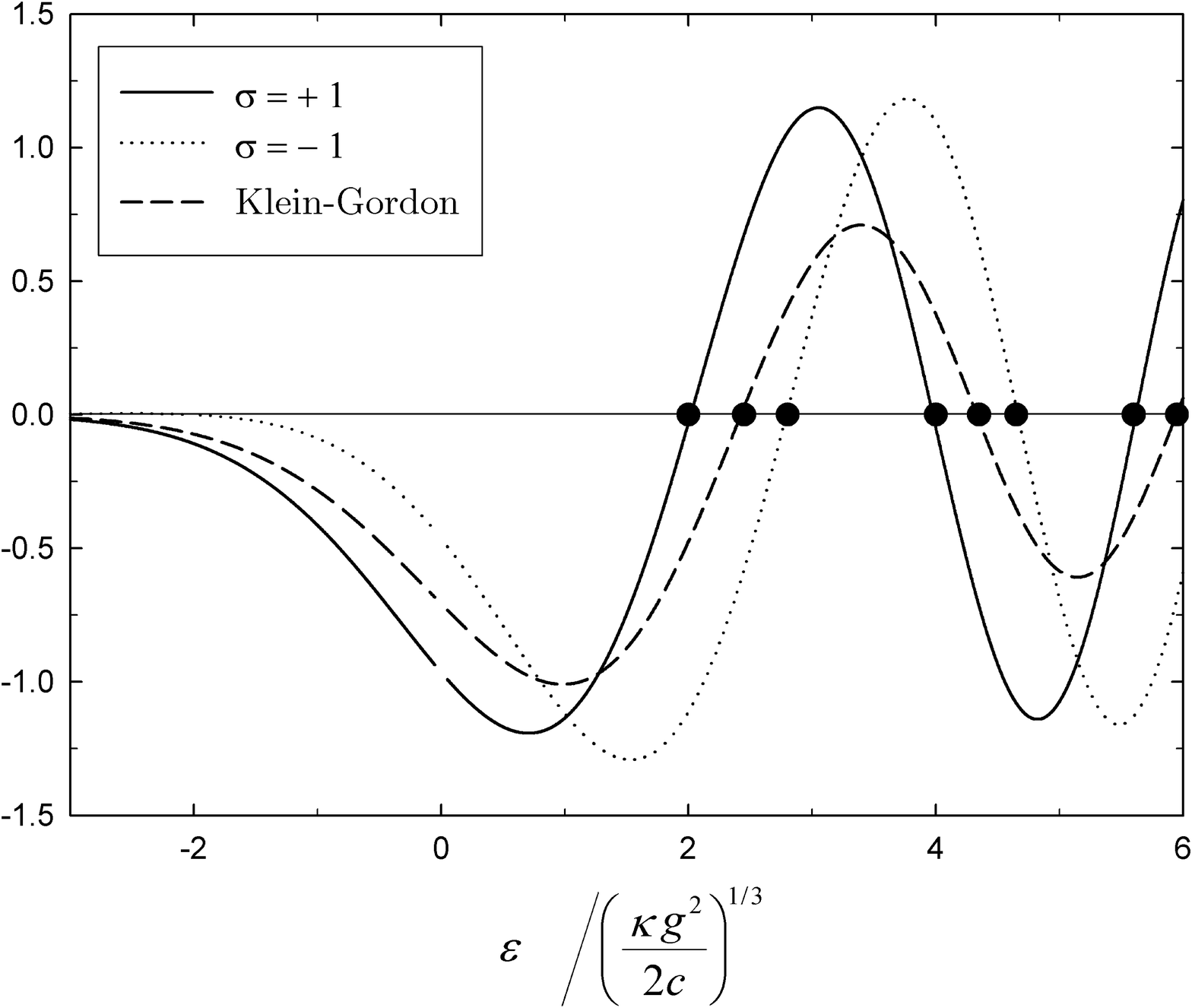}
\caption{Numerical plot of the left-hand side of the Dirac boundary condition~(\ref{bexa}) for $\sigma=+1$ (solid line) and $\sigma=-1$ (dotted line), and of the Klein-Gordon boundary condition~(\ref{bkg}) (dashed line). The zeros of these curves (full circles) are the energy eigenvalues, $\varepsilon_{n\sigma}$, given in units of $(\kappa\, g^2/2c)^{1/3}$. We fixed $\kappa=2,\,  g=1\,$ and $c=1$.}
\label{Fig1}
\end{figure}

\begin{figure}[t]
\includegraphics*[width=10cm]{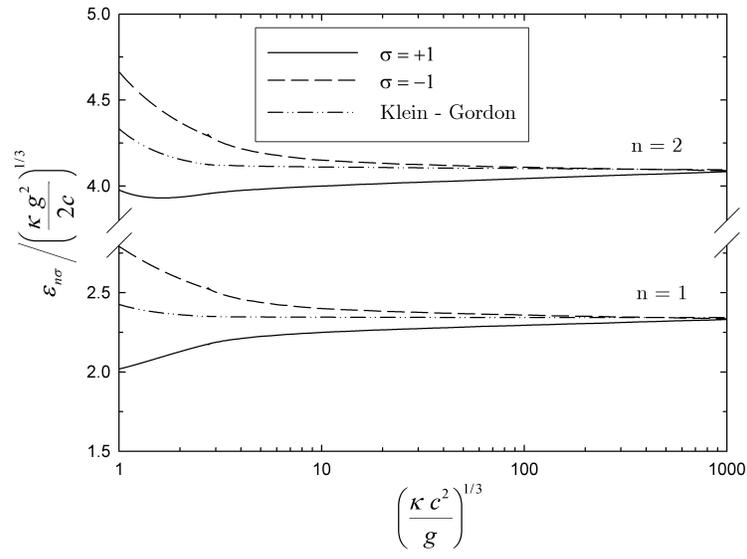}
\caption{Numerical plot of $\varepsilon_{n\sigma}$ in units of $(\kappa\, g^2/2c)^{1/3}$ versus the dimensionless parameter $(\kappa\, c^2/g)^{1/3}$ for the first two energy levels. We fixed $g=1$.}
\label{Fig2}
\end{figure}

 The spinor $\Theta_\sigma$ is an eigenspinor of the operator 
$\gamma^{\hat1}\gamma^{\hat2}\,$. 
Since this operator commutes with the Hamiltonian, $\sigma$ is a good quantum number. 
Depending on its values, the large and small components of $\Theta_\sigma$ can be identified 
with the functions $F$ and $G$ by the relations 
\begin{eqnarray}
	F\propto\lambda^- + \sigma\,\lambda^+\,, \quad G\propto\lambda^- -\sigma\, \lambda^+ \,.
\nonumber
\end{eqnarray}
The position of the mirror being given by $\rho=1/g$ and because 
$\lambda^-=(\lambda^+)^*$, the boundary condition~(\ref{quant0}) can be rewritten
\begin{equation}\label{bexa}
	\Re[H^{(1)}_{i\omega/g + 1/2}(i\kappa/g)]+\sigma\,\Im[H^{(1)}_{i\omega/g 
	+ 1/2}(i\kappa /g)]=0\,.
\end{equation}
Let us emphasize that if the spin contributions (the $1/2$ in the index of the Hankel functions) are neglected, these Hankel functions become real and Eq.~(\ref{bexa}) 
yields
\begin{equation}\label{bkg}
	H^{(1)}_{i\omega/g}(i\kappa/g)=0\,,
\end{equation}
which is the usual boundary condition in the Klein--Gordon case~\cite{KG2}. 
A plot of the left-hand side of the boundary conditions~(\ref{bexa}) compared to~(\ref{bkg}) 
is drawn in Fig.~\ref{Fig1}, where we defined $\varepsilon=\omega-\kappa\, c$, restoring the 
$c$ factors. 
Obviously, it leads to positive, quantized energy levels $\varepsilon_{n\sigma}\,$. 
In the nonrelativistic limit, these levels all coincide. This is illustrated in 
Fig.~\ref{Fig2}. 
\par In order to discuss this nonrelativistic limit, it is useful to introduce the
following quantities 
\begin{eqnarray}
	u=i\,\frac{\omega c}{g} + \frac{1}{2}\,,\quad w=\frac{i\kappa\, c^2}{gu}\, .
\nonumber
\end{eqnarray}
For weak gravitational fields, we have $|u|\gg 1\,$ and consequently $|w|\approx 1\,$. 
In this case, we can assume that ${\varepsilon}/\kappa\, c \ll 1$ and approximate the Hankel function 
by~\cite[Eq. (9.3.37)]{Abra} 
\begin{eqnarray}
	H^{(1)}_u(uw) &\propto& {\rm Ai}\left[-\left(\frac{2}{m g^2}\right)^{1/3}
\varepsilon\right]+\frac{i}{c}\left(\frac{g}{4m}\right)^{1/3}{\rm Ai}'\left[-\left(\frac{2}{m g^2}\right)^{1/3}
\varepsilon\right]+O(c^{-2})\, ,
\label{condiHank}
\end{eqnarray}
where ${\rm Ai}$ denotes the regular Airy function. Note that $\kappa\approx m\, c$ in the 
nonrelativtistic limit. Moreover, the imaginary part of $H^{(1)}$ can be neglected in this 
limit, so that the condition~(\ref{bexa}) together with Eq.~(\ref{condiHank}) gives 
\begin{eqnarray}\label{nonrel}
	\varepsilon_n	&\approx&-\left(\frac{m g^2}{2}\right)^{1/3}\alpha_n\;,
\end{eqnarray}
with $\alpha_n$ the $n^{{\rm th}}$ zero of the regular Airy function. 
These can be found for example in Ref. \cite[Table 10.13]{Abra}. 
A WKB approximation of these zeros can also be found in \cite[Eq. (10.4.94)]{Abra}:   
\begin{equation}\label{dh1}
\alpha_n=-\left[\frac{3\pi}{2}(n-1/4)\right]^{2/3}\,.
\end{equation}
This approximation is precise up to $8\%$. Note that the asymptotic values of the energy levels $\varepsilon_{n\sigma}$ in Fig.~\ref{Fig2} are given by $-\alpha_1\approx2.34$ and $-\alpha_2\approx4.09$ as expected from Eq.~(\ref{nonrel}).

\subsection{Relativistic expansion}\label{BsKR2}
 
%
\subsubsection{Lowest order solutions}\label{Solu}
%

Although it is possible to find an analytical solution of the Dirac equation in Rindler 
spacetime, 
the determination of the energy is rather problematic in the other cases, due to the 
nature of the singularity in the differential equations. 
To our knowledge, no explicit solution in Kasner--Taub geometry has been found yet. 
It is thus interesting to expand the equations in powers of $1/c^2\,$. 
Moreover, this procedure will allow us to find a solution in Kasner--Taub spacetime and to study 
the influence of a weak magnetic field. The equation we need to solve, that is Eq. (\ref{Dirstep4}), 
can be expanded up to the order $c^{-2}$ with the appropriate choice of $A$ and $B$ given by  
Eqs.~(\ref{defc2}). We define the energy $\omega$ as 
\begin{equation}\label{omegadef}
\omega=mc^2+\frac{k^2}{2m}+{\cal E}.	
\end{equation}
>From Eq.~(\ref{Dirstep4}), we obtain at lowest order 
\begin{equation}\label{ham0}
	-\frac{f''}{2m}+mg\zeta f-\frac{\mu_n}{2}\,{\cal B}\,\sigma f={\cal E}f\,,
\end{equation}
which is the expected Pauli equation with a linear gravitational potential and a magnetic 
field. To go ahead, we simply assume that the magnetic field is constant in the $y$ direction, \textit{i.e.}
\begin{equation}
	{\cal B}={\cal B}_0\, .
\end{equation}
This choice is relevant in the context of the GRANIT experiment~\cite{nesv02} since it can 
easily be added to the current experimental setup. Equation (\ref{ham0}) is then simply a 
Schr\"odinger equation with a linear potential, with the boundary condition~(\ref{quant}) given 
at this order by $\left.f\right|_{\zeta=0}=0\,$. 
The solution of such an eigenequation is well-known \cite[Problem 40]{Flu} 
\begin{subequations}\label{solunr}
\begin{equation}\label{en0}
{\cal	E}_{n\sigma}=-\left(\frac{mg}{\theta}\,\right) 
\alpha_n-\frac{\mu_n}{2}\,{\cal B}_0\,\sigma\, ,
\end{equation}
\begin{equation}\label{fo_un}
	f_{n\sigma}=\frac{\theta^{1/2}}{|{\rm Ai}'(\alpha_n)|}
	 {\rm Ai} \left[\theta\, \zeta+\alpha_n\right]\,,
\end{equation}
with 
\begin{eqnarray}\label{thetadef}
	\theta=\left(2m^2 g\right)^{1/3}\,.
\end{eqnarray}
\end{subequations}
Without magnetic field, Formula (\ref{en0}) tells us that the energy of the particle, and thus the height at which it can bounce, only depends on its mass and of the strength of the gravitational field. If a nonzero magnetic field is present, the energy depends on $\sigma$. As it is deduced from Eqs.~(\ref{u_prop}), $\sigma/2$ can be interpreted as the spin of the particle along the $y$ direction. This will be detailed in the following.  
  
As a consistency check, we note that the formula (\ref{energy}), taken in the
nonrelativistic limit where $F=f$ and $G=0\,$, gives
\begin{equation}\label{omega0}
	\omega_0=mc^2+\frac{k^2}{2m}+\frac{\left\langle p^2_\zeta\right\rangle}{2m}+mg\left\langle \zeta\right\rangle-\frac{\mu_n\sigma}{2}\left\langle {\cal B}\right\rangle.
\end{equation}
The average values are computed with respect to the wave function~(\ref{fo_un}). 
The expressions for these average values are analytical, 
as it can be seen in Eqs.~(\ref{prop}). 

\subsubsection{Relativistic corrections}
\par We are mainly interested here in the relativistic corrections to the energy spectrum. These can be obtained by expanding the formula (\ref{energy}) at the order $c^{-2}\,$. 
The wave functions which have to be used are the properly normalized $F$ and $G$ computed 
thanks to Relations~(\ref{norm}), (\ref{Gdef}) and (\ref{Fdef}), with $f$ given by (\ref{fo_un}). We can keep the same boundary conditions as in previous section. After some algebra, we find  
\begin{eqnarray}
	\omega=\omega_0+\omega_2+O(c^{-4})\,,
\nonumber
\end{eqnarray}
with $\omega_0$ the nonrelativistic energy (\ref{omega0}) and
\begin{equation}\label{pertu}
	\omega_2=-\frac{\left(k^2+\left\langle p^2_\zeta\right\rangle\right)^2}{8m^3c^2}+\left[\frac{\left\langle\zeta p^2_\zeta\right\rangle}{2m}+(1+2b)
	\frac{\left\langle \zeta\right\rangle k^2}{2m}\right]\frac{g}{c^2}
	+(1-a)\frac{g^2 m \left\langle \zeta^2\right\rangle }{c^2}+(1+2b)\frac{gk\sigma}{4mc^2}\,.
\end{equation}
\par The various relativistic corrections appearing in the above equation are now interpreted in analogy with the energy 
$E$ of a classical particle on a curved spacetime. 
With the metric~(\ref{met_ansa}), we obtain
\begin{eqnarray}
	E=A\sqrt{m^2c^4+B^{-2}c^2k^2+c^2p^2_\zeta}
\nonumber
\end{eqnarray}
and thus
\begin{eqnarray}\label{ecla1}
	E&\approx& mc^2+\frac{k^2}{2m}+\frac{p^2_\zeta}{2m}+mg \zeta\nonumber\\
	&&-\frac{(k^2+p^2_\zeta)^2}{8m^3c^2}+\frac{g \zeta}{c^2}\frac{p^2_\zeta}{2m}+(1+2b)\frac{g \zeta}{c^2}\frac{k^2}{2m}
	+(1-a)m\frac{(g\zeta)^2}{c^2}
	\,.
\end{eqnarray}
Apart from the spin ($\sigma$) dependent term, Eq.~(\ref{pertu}) clearly reproduces the terms appearing in the classical formula~ (\ref{ecla1}). The first term in $\omega_2$ is the usual relativistic correction arising from the expansion of the relativistic kinetic energy in powers of $1/c^2$. The next two terms between square brackets are redshift corrections. 
The first one gives a quantum mechanical translation of the usual redshift formula for a particle falling in a gravitational field,
\textit{viz}
\begin{eqnarray}
	E(L)=\left(1+\frac{gL}{c^2}\right)E(0)\,.
\nonumber
\end{eqnarray}
The second is an additional transverse redshift term. The $\zeta^2$ term is a curvature correction. As expected, it vanishes in the flat Rindler geometry. The last term is a spin-dependent correction.
\par The properties of the regular Airy function lead to the relations  
\begin{subequations}\label{prop}
\begin{equation}
	\langle \zeta\rangle=-\frac{2\, \alpha_n}{3\, \theta},\quad \langle \zeta^2\rangle=\frac{8\, \alpha^2_n}{15\, \theta^2},
\end{equation}
\begin{equation}
	\left\langle p^2_\zeta\right\rangle=-\theta^3\left\langle \zeta\right\rangle-\theta^2\alpha_n,\quad \left\langle \zeta p^2_\zeta\right\rangle=-\theta^3\left\langle \zeta^2\right\rangle-\theta^2\alpha_n \left\langle \zeta\right\rangle,
\end{equation}
\end{subequations}
which allow to express Eq.~(\ref{pertu}) in terms of $\alpha_n$~(\ref{dh1}) and $\theta$ (\ref{thetadef}). 

\section{Foldy-Wouthuysen transformation}\label{Pauli}

%
\subsection{Kasner--Taub and Rindler metrics}\label{KRFW}
%
It is interesting to compare the previous energy spectra with the results obtained by resorting to a FW transformation~\cite{BandD,FW}. 
Whether this procedure works in the case we are dealing with is indeed still subject to discussions~\cite{Obu,Sil}. 
\par The Dirac equation~(\ref{Direxpl}) can be recast in the Hamiltonian form
\begin{equation}\label{Dire3}
	i\partial_t\Psi=H\Psi,
\end{equation}
where 
\begin{eqnarray}
\label{ham1}
	H&=&\beta m - \mu_n\beta\Sigma_{\hat{k}}{\cal B}^{\hat{k}}+{\cal O}+{\cal P},\\
	{\cal O}&=&AB^{-1} \alpha^{\hat{\jmath}} p_j+A\alpha^{\hat3} p_\zeta-\frac{i}{2}\left(A'+\frac{2B'A}{B}\right)\alpha^{\hat3}\,,
	\quad j=1,2\,,
	\nonumber \\
	{\cal P}&=&m(A-1)\beta\,.
\nonumber
\end{eqnarray}
We recall that the operators ${\cal O}$ and ${\cal P}$ satisfy
\begin{eqnarray}
	\left\{\beta,{\cal O}\right\}=0=\left[\beta,{\cal P}\right]
\nonumber
\end{eqnarray}
and that $p_k=-i\partial_k\,$.  Using the standard Bjorken and Drell conventions, at the first order in $1/m$, the FW Hamiltonian computed from~(\ref{ham1}) is \cite{BandD}
\begin{eqnarray}
	H_{{\rm FW}}=\beta m+\beta\frac{{\cal O}^2}{2m}+{\cal P}
	-\frac{1}{8m^2}\left[{\cal O},\left[{\cal O},{\cal P}\right]\right]\,.
\nonumber
\end{eqnarray}
The positive-energy part of this FW Hamiltonian is given by
\begin{eqnarray}\label{FW_rind}
	H_{{\rm FW}}&=&mc^2+\left[1+(1+2b)
	\frac{\vec g\cdot\vec x}{c^2}\right]\frac{k^2}{2m}
	+\left[1+\frac{\vec g\cdot\vec x}{c^2}\right]\frac{p^2_\zeta}{2m}
	+ m\vec g\cdot \vec x
	-\mu_n \vec  S\cdot\vec{\cal B}
\nonumber\\
	&&+(1+2b)\frac{\vec S\cdot (\vec g\times\vec p)}{2mc^2}
	+(1-a)m\frac{(\vec g\cdot\vec x)^2}{c^2}
	+i(2b-1)\frac{\vec g \cdot\vec p}{2mc^2}\,,
\end{eqnarray}
with $\vec g=(0,0,g)$. 
The lowest order FW Hamiltonian is almost identical to the one which can be read from Eq.~(\ref{ham0}) provided that one sets $S_y=\sigma/2\, $. This confirms the interpretation of $\sigma/2$ given in the previous sections. 
Moreover, the relativistic corrections are on average identical to those obtained in the formula~
(\ref{pertu}) since $\left\langle p_\zeta\right\rangle=0$. As the FW transformation is only performed up to the order $m^{-1}\,$, we miss the kinetic corrections in $m^{-3}$. 
So, comparing the method used in the previous section with the FW method  
at the same order in $1/m$, 
we may conclude that both approaches lead to the same results. 
Let us also note that for the Rindler metric without magnetic field, we recover the result 
of Ref.~\cite{Hehl}.

%
\subsection{Schwarzschild metric}\label{SW}
%

Since we have shown that the usual relativistic expansion agrees with the FW technique for Kasner--Taub and Rindler geometries, we apply the latter to the Schwarzschild geometry. The Hamiltonian corresponding to Eq.~(\ref{Dir_sch}) is 
\begin{eqnarray}
	H=-i\alpha^{\hat3}{\rm e}^{\Phi-\Lambda}
	\left(\partial_r+\frac{\Phi'}{r}+\frac{1}{r}\right)
	-i\alpha^{\hat1}\frac{{\rm e}^\Phi}{r}\left(\partial_\theta
	+\frac{\cot\theta}{2}\,\right)
	-i\alpha^{\hat2}\frac{{\rm e}^\Phi}{r\sin\theta}\partial_\phi
	+m{\rm e}^\Phi\beta
	\,.
\nonumber
\end{eqnarray}
After a FW transformation, we find that the positive-energy part of the Hamiltonian is 
\begin{eqnarray}\label{fw_sch}
	H_{{\rm FW}}&=&mc^2+\left[
	1+\frac{3\vec g \cdot \vec r}{c^2}\,\right]\frac{p^2_r}{2m}\,+
	\left[1+\frac{\vec g\cdot \vec r}{c^2}\,\right]\frac{\ell(\ell+1)}{2mr^2}\,
	+ m\,\vec g\cdot \vec r
	\nonumber\\
	&&-\frac{3}{2mc^2}\,\vec S \cdot (\vec g\times\vec p)-\frac{m}{2}\,
	\frac{(\vec g \cdot \vec r)^2}{c^2}\,+i\,\frac{\vec g \cdot \vec p}{mc^2}\; ,
\end{eqnarray}
where $\vec g=-GM\vec 1_r/r^2\,$ and $\ell(\ell+1)$ is the eigenvalue of the squared orbital momentum. 
Formula~(\ref{fw_sch}) agrees with the corresponding formula (28) of Ref.~\cite{Sil} upon performing a change of coordinates 
from isotropic to the usual Schwarzschild coordinates. Here again, $\left\langle p_r\right\rangle=0$ and the last term in Hamiltonian~(\ref{fw_sch}) brings no contribution.

Let us point out that the classical energy of a nonrelativistic particle moving in the spacetime (\ref{met_sphe})
geometry is given by 
\begin{eqnarray}
	E=  {\rm e}^{\Phi}\sqrt{m^2c^4 +{\rm e}^{-2\Lambda}c^2p^2_r+c^2\frac{L^2}{r^2}}\,.
\nonumber
\end{eqnarray} 
In the particular case of the Schwarzschild metric, we have
\begin{eqnarray}\label{ecla2}
	E&\approx& mc^2+\frac{p^2_r}{2m}\,+\frac{L^2}{2mr^2}\,+ m\vec g\cdot \vec r\nonumber\\
	&&-\frac{(p^2_r+L^2/r^2)^2}{8m^3c^2}\,
	+3\,\frac{\vec g \cdot \vec r}{c^2}\,\frac{p^2_r}{2m}\,
	+\frac{\vec g\cdot \vec r}{c^2}\,\frac{L^2}{2mr^2}\,
	-\frac{m}{2}\,\frac{(\vec g \cdot \vec r)^2}{c^2}\, . 
\end{eqnarray}
The classical energy given by Eq.~(\ref{ecla2}) is again identical to the Hamiltonian~(\ref{fw_sch}) apart from the spin-dependent terms and the kinetic corrections in $m^{-3}$. 
\section{Comparison with the GRANIT experiment}\label{compar}

\par What can be measured experimentally is the critical height, corresponding to the classical turning point of the neutron. It is given by
\begin{equation}
	{\cal E}_{n\sigma}=m g h_{n\sigma}, 
\end{equation}
 which can be rewritten in our approximations
\begin{equation}\label{exp_h}
	h_{n\sigma}=-\frac{\alpha_n}{\theta}-\frac{\mu_n{\cal B}_0\sigma}{2mg}+\frac{\omega_{2,n\sigma}}{m g}\, .
\end{equation}
At the lowest order, and without magnetic field, we simply obtain
\begin{equation}
	h_n=-\frac{\alpha_n}{\theta}\, .
\end{equation}
This relation has been successfully checked for the first two bound states by the GRANIT experiment \cite{nesv02}. Indeed, knowing that $\theta=0.17\, \mu$m$^{-1}$ for a neutron in the Earth's gravitational field, the predicted heights are  
\begin{eqnarray}\label{engr}
	h_1&=&13.7\  \mu{\rm m}, \quad\quad h_2=24.0\  \mu{\rm m}\,,
\end{eqnarray}
while the experimental results concerning these states are
\begin{equation}
	h^{{\rm exp}}_1=12.2\ \mu{\rm m}\pm1.8_{{\rm syst}}\pm0.7_{{\rm stat}}\,, \quad
	h^{{\rm exp}}_2=21.6\ \mu{\rm m}\pm2.2_{{\rm syst}}\pm0.7_{{\rm stat}}\,.
\end{equation}

\par If we still stay at the lowest order but allow for a weak magnetic field, then, formula~(\ref{exp_h}) predicts a splitting of the critical heights depending on the values of $\sigma$. A given level $h_n$ shall be split in $h_{n,1}$ and $h_{n,-1}\, $, both levels being separated by a value which does not depend on $n$, that is
\begin{equation}
	\Delta h\approx\left|\frac{\mu_n {\cal B}_0}{mg}\right|\,.
\end{equation}
In the future of the GRANIT experiment, it seems possible to deal with a magnetic field whose density $\nu$ is of order $0.1\, T/m$~\cite{private}. As the characteristic size of the experiment is around $10\, \mu$m, a value such as ${\cal B}_0\approx 10^{-6}$ T could be produced (we assume the experimental apparatus to be shielded from the Earth's magnetic field). In such circumstances, one would observe $\Delta h\approx\, 0.6\, \mu$m. An experimental confirmation of this point requires an increase of the experimental accuracy, which should be possible in a near future~\cite{private}. Note that a value such as $\nu=0.1\, T/m$ is still in the weak magnetic field regime since $\mu_n\nu/mg\approx6\, \%$.
\par The relativistic corrections are particularly interesting, since they involve couplings between spin, momentum, and gravity. In particular, there exists the term $\vec S\cdot (\vec g\times\vec p)$, which states that even without magnetic field, the particle will bounce at different heights depending on the value of its spin. Again, this term will split a given height-level into two separate levels, distant by the quantity 
\begin{equation}
	\Delta h_{sg}\propto\frac{\hbar \,v_h}{2mc^2}\,,
\end{equation}
where $v_h$ is the horizontal speed of the particle. In the GRANIT experiment, $v_h\approx 6.5$~m/s~\cite{nesv02}. This leads to $\Delta h_{sg}=O(10^{-18}\mu$m$)$: The experimental detection of this phenomenon unfortunately seems hopeless with such an experiment, as well as the detection of the other relativistic corrections, as pointed out in Ref.~\cite{Arm}. 

\section{Conclusion}\label{Conclu}
In this work, we obtained the quantization of the energy of a Dirac particle bouncing on a mirror on a curved spacetime. Consequently, the critical heights at which the particle 
bounces are also quantized. This is due to the presence of the mirror, which imposes a Robin boundary 
condition leading to the existence of a discrete set of bound states. We computed the spectrum of these bound 
states for several choices of spacetime metrics. 
Firstly, we solved the problem analytically on the Rindler spacetime. We then computed the energy spectrum in Kasner--Taub and Rindler geometries at the second order in 
$1/c^2$ thanks to a relativistic expansion. At lowest order, these energies are 
equal in both geometries - in agreement with the equivalence principle. However, they differ in the relativistic corrections through redshift, curvature and spin terms. One of the most conceptually interesting corrections 
is a term in $\vec S\cdot (\vec g \times\vec p)$, 
which shows that the energies, and thus the critical heights, are spin-dependent. 
We also expressed the Hamiltonian in Rindler, Kasner--Taub, and Schwarzschild geometries using 
a Foldy--Wouthuysen transformation. We checked that the results are identical for the first two geometries, 
showing \textit{a posteriori} that the Foldy--Wouthuysen transformation is valid in this framework. Finally, we compared our results with those of the GRANIT experiment. 
The nonrelativistic approximation is enough the reproduce the current experimental results, but the relativistic corrections appear too be small to be detected. Moreover, we showed that the presence of a weak constant magnetic field leads to observable effects. 

\acknowledgments
One of us (Ph.S.) would like to thank F. Englert and R. Parentani for illuminating discussions. The work of N.B. and F.B. is supported by the FNRS (Belgium). This work is supported in part by IISN-Belgium (convention 4.4511.06).

\begin{thebibliography}{99}
\bibitem{hawk} We don't mention the celebrated Hawking effect which concerns 
second-quantized systems.
\bibitem{fn2}A maybe more satisfactory boundary
condition could be obtained by applying to the Dirac field the model of Ref.~\cite{Par}
developed for a scalar field. However, this would imply a second quantization, which is
not considered in the present paper.
\end{thebibliography}
\end{document}